\documentclass[aps,prl,reprint,groupedaddress,showpacs]{revtex4-1}

\usepackage{graphicx}
\usepackage{dcolumn}
\usepackage{bm}

\usepackage{amsmath} 
\usepackage{amsfonts} 
 
\begin{document}


\title{Evolution of the Kondo Effect in a Quantum Dot Probed by the Shot
Noise}

\author{Yoshiaki Yamauchi$^1$, Koji Sekiguchi$^1$, Kensaku Chida$^1$,
Tomonori Arakawa$^1$, Shuji Nakamura$^1$, Kensuke Kobayashi$^1$}
\email{kensuke@scl.kyoto-u.ac.jp}

\author{Teruo Ono$^1$, Tatsuya Fujii$^2$, Rui Sakano$^3$}

\affiliation{$^1$Institute for Chemical Research, Kyoto University, Uji,
Kyoto 611-0011, Japan.}

\affiliation{$^2$Institute for Solid State Physics, University of Tokyo,
Kashiwa, Chiba 277-8581, Japan.}

\affiliation{$^3$Department of Applied Physics, University of Tokyo,
Hongo, Tokyo 113-0033, Japan.}

\begin{abstract}
We measure the current and shot noise in a quantum dot in the Kondo
regime to address the nonequilibrium properties of the Kondo effect. By
systematically tuning the temperature and gate voltages to define the
level positions in the quantum dot, we observe an enhancement of the
shot noise as temperature decreases below the Kondo temperature, which
indicates that the two-particle scattering process grows as the Kondo
state evolves. Below the Kondo temperature, the Fano factor defined at
finite temperature is found to exceed the expected value of unity from
the noninteracting model, reaching $1.8 \pm 0.2$.
\end{abstract}

\date{\today}


\maketitle 

As the Kondo effect is one of the most fundamental many-body phenomena
in condensed matter physics~\cite{KondoPTP1964}, its realization in a
quantum dot (QD)~\cite{Goldhaber-GordonNature1998} would be an
attractive stage at which various theoretical predictions for Kondo
physics can be tested in a way otherwise impossible.  Indeed, the
dependence of this effect on several external parameters such as
temperature, bare level position, and magnetic field has been precisely
compared with
theory~\cite{Goldhaber-GordonNature1998,vanderWielScience2000}. Many
other aspects of the Kondo state, such as its
coherence~\cite{vanderWielScience2000,JiScience2000,SatoPRL2005}, have
been tested, which has deepened our understanding of how a local spin
interacts with continuum to form a correlated ground state.

Most of the experiments performed thus far using a Kondo QD have focused
on the equilibrium properties of the Kondo effect; however, it is also
possible to investigate the nonequilibrium correlated states (for
example, see Refs.~\cite{FranceschiPRL2002,GrobisPRL2008}). An
experiment on the shot noise or the nonequilibrium fluctuation in the
current~\cite{BlanterPR2000} passing through a Kondo
QD~\cite{ZarchinPRB2008,DelattreNatPhys2009} would be valuable as it
would yield quantitative values for comparison with existing
theories~\cite{MeirPRL2002,SelaPRL2006,GolubPRB2006,GogolinPRL2006,MoraPRL2008,MoraPRB2009_2,FujiiJPSJ2010,SelaPRB2009}.
Several theories predict that shot noise in a Kondo QD is enhanced by
two-particle scattering with a Fano factor larger than that expected in
a noninteracting case. A pioneering experiment was performed to tackle
this problem~\cite{ZarchinPRB2008}; however, the versatile
controllability of a Kondo QD and the universality of Kondo physics have
still not been investigated in detail.

Here, we report an experiment for the measurement of current and shot
noise through a Kondo QD. Upon systematic tuning of the temperature and
gate voltages, we observed an enhancement of the shot noise in the QD
with a decrease in the temperature to below the Kondo temperature, which
indicates the evolution of the nonequilibrium Kondo state.  We discuss
this observation in terms of the finite-temperature Fano
factor~\cite{MoraPRL2008,MoraPRB2009_2,FujiiJPSJ2010,FujiiJPSJ2007},
which, well below the Kondo temperature, exceeds the expected value of
unity from the noninteracting model.

The experiment was performed on a QD fabricated on GaAs/AlGaAs
two-dimensional electron gas (electron mobility of $2.7 \times 10^5$
cm$^2/$Vs and electron density of $2.4 \times 10^{11}$ cm$^{-2}$ at 4.2
K) with four Au/Ti metallic gate electrodes patterned by electron beam
lithography. A scanning electron microscopy (SEM) image of the QD and a
schematic of the experimental setup for measuring the current and noise
in the dilution refrigerator are shown in Fig.~1(a).

The noise was measured as
follows~\cite{de-PicciottoNature1997,DiCarloRSI2006,HashisakaPRB2008}. The
voltage fluctuation $S_v$ across the sample on the resonant ($LC$)
circuit, whose resonance frequency was set to 2.65~MHz with a bandwidth
of $\sim 140$~kHz, was extracted as an output signal of the cryogenic
amplifier followed by a room-temperature amplifier. The time-domain
signal was captured by a digitizer and converted to spectral density via
fast-Fourier transform. Fitting of the resonance peak gave $S_v$. To
derive the current noise power spectral density $S_i$ of the sample, the
noise measurement setup had to be calibrated.  As the cryogenic
amplifier has finite voltage and current noises ($S_v^{amp}$ and
$S_i^{amp}$ in terms of the amplifier input), the measured $S_v$ for the
sample with resistance $R$ is related to $S_i$ as $S_v = A
\left(S_v^{amp} + 1/(1/R+1/Z)^2(S_i^{amp}+S_i) \right)$. Here, $A$ and
$Z$ are the gain of the amplifiers and the impedance of the $LC$ circuit
at the resonance frequency, respectively. To extract these parameters,
the thermal noise $S_i = 4 k_BT/R$ (where $k_B$ is the Boltzmann
constant) was measured for about 30 different $R$'s between 10~k$\Omega$
and 70~k$\Omega$ at eight different temperatures ($T$) below 800 mK.  We
obtained $A = 1.1 \times 10^6$~V$^2/$V$^2$, $Z=72$~k$\Omega$,
$S_v^{amp}= 8.4 \times 10^{-20}$~V$^2/$Hz, and $S_i^{amp}= 1.9 \times
10^{-28}$~A$^2/$Hz. To derive the current noise at the QD ($S_I$) from
$S_i$, the thermal noise of the contact resistance was taken into
account. We confirmed that the electron (noise) temperature $T$ can be
controlled to be between 130 mK and 800 mK. The estimated resolution of
$S_I$ of the present setup was $2 \times 10^{-29}$~A$^2/$Hz.

\begin{figure}[tbp]
\center \includegraphics[width=.87\linewidth]{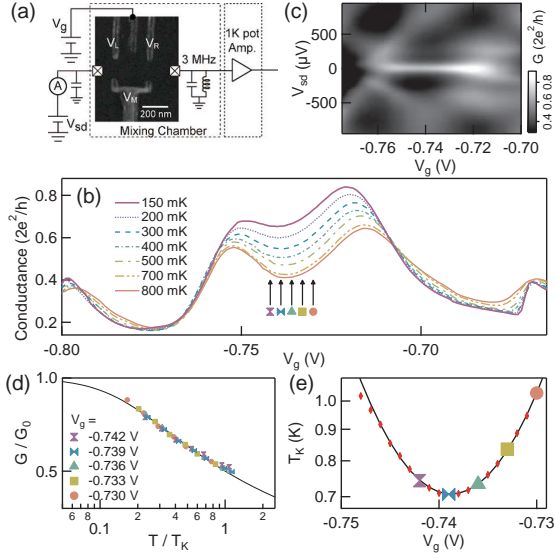}
\caption{ (a) SEM image of a QD in the current and noise
measurement setup in a dilution refrigerator. The negative voltages
$V_g$, $V_L$, $V_R$, and $V_M$ are applied to the gates to define the
QD. (b) Conductance as a function of $V_g$ at various temperatures
between 150 mK and 800 mK. The positions for $V_g = -0.742$, $-0.739$,
$-0.736$, $-0.733$, and $-0.730$ V are indicated by special marks (see
(d) and (e)). (c) Coulomb diamond at 150 mK with a Kondo resonance
at around $V_g \sim -0.74$~V. (d) Scaling plot of conductance as a
function of $T$. (e) Gate voltage dependence of $T_K$. The curve
represents the expected $T_K$ for $U = 0.56$~meV and $\Gamma = 0.34$~meV.}
\end{figure}

Upon tuning the voltages of the four gate electrodes ($V_g$, $V_L$,
$V_R$, and $V_M$ shown in Fig.~1(a)), the conductance $G$ through the QD
showed typical behavior expected for the Kondo effect. Further, as shown
in Fig.~1(b), upon decreasing $T$ from 800 mK to 150 mK, the conductance
at the Coulomb valley centered around $V_g \sim -0.74$~V enhanced,
whereas conductance in the neighboring valleys around $V_g \sim -0.77$~V
and $-0.68$~V did not show enhancement. This indicates that the number
of electrons is odd at the valley around $V_g \sim -0.74$~V and the
Kondo effect appears at low
temperatures~\cite{Goldhaber-GordonNature1998}. In fact, the
differential conductance at finite source-drain voltages ($V_{sd}$) at
$T = 150$~mK shows a peak structure, namely, Kondo resonance, between
$V_g = -0.71$~V and $-0.75$~V as shown in Fig.~1(c).

To quantitatively discuss the Kondo effect in a QD, it is important to
derive several basic parameters to characterize it. To this end, we
adopted a previously reported method
\cite{vanderWielScience2000,SatoPRL2005,GrobisPRL2008,Goldhaber-GordonPRL1998}.
First, we numerically fit the temperature dependence of the conductance
at a given $V_g$ to derive the $V_g$-dependent Kondo temperature
($T_K$), as $V_g$ modulates the bare level position and hence $T_K$. We
used the empirical function $G(T) = G_0 /(1+(2^{1/s}-1)(T/T_K)^2)^s$,
where $s \approx 0.2$ for a spin-$1/2$ Kondo system. By taking $s=0.18$
and $G_0 = 0.83 G_q$ ($G_q \equiv 2e^2/h \sim (12.9$~k$\Omega)^{-1}$),
we adequately scaled the temperature dependence of the conductance as
shown in Fig.~1(d), where the normalized conductance $G/G_0$ is plotted
against the normalized temperature $T/T_K$. As the two barriers for
defining the QD are asymmetric~\cite{Goldhaber-GordonPRL1998}, $G_0/G_q$
is less than unity.  The obtained $T_K$'s are plotted as a function of
$V_g$ in Fig.~1(e). $T_K$ was determined by the charging energy ($U$),
the energy level of the single-particle state ($\epsilon_0$), and its
width ($\Gamma$) as $T_K = \sqrt{\Gamma U}/2 \exp ((\pi
\epsilon_0(\epsilon_0+U)/\Gamma U)$.  By numerical fitting, $U$ and
$\Gamma$ were obtained as 0.56 meV and 0.34~meV, respectively.  The
obtained value of $U$ is consistent with the Coulomb diamond shown in
Fig.~1(c).

\begin{figure}[tbp]
\center \includegraphics[width=.85\linewidth]{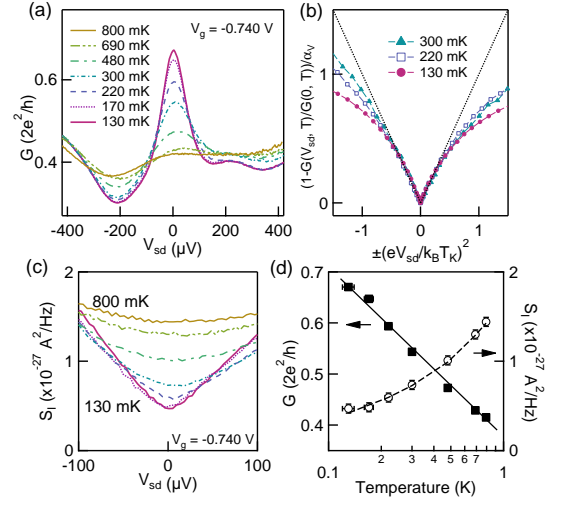}
\caption{(a) Differential conductance $G(V_{sd}, T)$ at $V_g = -0.740$~V
for several temperatures between 130~mK and 800~mK.
(b) Plot of $(1-G(V_{sd}, T)/G(0, T))/\alpha_V$ versus $(eV_{sd}/k_B T_K)^2$~\cite{GrobisPRL2008}. The solid
line shows the associated universal curve. (c) Current noise spectral
density $S_I (V_{sd})$ corresponding to $G (V_{sd})$ shown in (a). (d)
Temperature dependence of zero-bias conductance and thermal
noise at $V_g = -0.740$~V. }
\end{figure}

The differential conductance and current noise at $V_g = -0.740$~V,
which is close to the electron-hole symmetry point with the lowest $T_K$
of 0.70~K, are shown in Figs.~2(a) and (c), respectively. Since $\Gamma
\gg k_B T_K$, the low-energy properties are characterized by $k_B
T_K$. The differential conductance $G(T, V_{sd})$ at various $T$ between
130 mK and 800 mK is shown in Fig.~2(a). As $T$ decreases, a resonant
peak emerges at zero bias voltage. Figure~2(b) shows the scaling plot of
$G(T, V_{sd})$, where $(1-G(V_{sd}, T)/G(0, T))/\alpha_V$ is plotted as
a function of $(eV_{sd}/k_B T_K)^2$ with $\alpha_V = c_T \alpha
/(1+c_T(\gamma/\alpha-1)(T/T_K)^2)$ ($e>0$ is the electron
charge). Here, we took $\alpha = 0.10$, $\gamma = 0.50$, and $c_T =
\sqrt{2^{1/s}-1}$ as reported before~\cite{GrobisPRL2008} and achieved
good scaling of up to $\sim 0.5(eV_{sd}/k_B T_K)^2$, which supports the
validity of the present analysis.  Figure~2(d) shows the characteristic
logarithmic enhancement of the conductance with decreasing
temperature. Here, we superpose the equilibrium current noise (thermal
noise) as a function of $T$, which is in good agreement with the
Johnson-Nyquist relation $S_I = 4k_B T G(V_{sd}=0)$ represented by the
dashed curve.

Based on the above properties of our Kondo QD, we analyzed the current
noise $S_I$ according to a recently proposed
scheme~\cite{MoraPRL2008,MoraPRB2009_2,FujiiJPSJ2010,FujiiJPSJ2007} in
terms of the finite-temperature Fano factor for the Kondo system. The
finite-temperature Fano factor $F_K$ is defined as
\begin{equation}
F_K \equiv \frac{S_K}{2eI_K},
\end{equation}
where $I_K$ and $S_K$ are expressed using the observed current
$I(V_{sd}, T)$ and $S_I(V_{sd}, T)$ as $ I_K \equiv I_0 -I(V_{sd}, T)$
and $S_K \equiv S_I(V_{sd}, T) - 4 k_B T G(V_{sd}, T) - S_0$,
respectively. Here, $I_0 \equiv G_q (1-\delta^2)V_{sd}$, where
$\delta^2$ denotes the asymmetry of the two barriers of the QD. It
should be noted that $I_K$ is not the current back-scattered at the QD
as was the case in Ref.~\cite{ZarchinPRB2008}, but is rather the
deviation of the current from that at the Kondo fixed point.  Similarly,
at the fixed point, the asymmetry of the QD yields the corresponding
shot noise contribution as expressed by $S_0 \equiv 2eG_q \delta^2
(1-\delta^2)V_{sd}(\coth(x)-1/x)$ with $x \equiv eV_{sd}/2k_B T$. In the
absence of the Kondo correlation with $U/\Gamma \rightarrow 0$, $F_K =
1$, which in agreement with the noninteracting
case~\cite{FujiiJPSJ2010,SelaPRB2009,SakanoFujiiOguri}.

\begin{figure}[tbp]
\center \includegraphics[width=.75\linewidth]{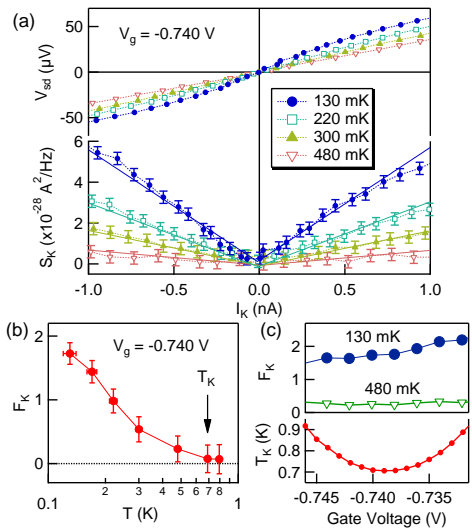}
\caption{(a) (upper) Relation between $V_{sd}$ and $I_K$ plotted for
$V_g = -0.740$~V and at various temperatures. (lower) Corresponding plot
showing $S_K$ as a function of $I_K$. The solid curves show the result
of numerical fitting to obtain $F_K$. (b) $F_K$ as a function of
temperature. (c) Gate voltage dependence of $F_K$ with $T_K$ as a
function of $V_g$.}
\end{figure}

Figure~3(a) shows the result of analysis of the current and shot noise
obtained at $V_g = -0.740$~V for 130, 220, 300, and 480 mK. The upper
panel shows the relation between $I_K$ and $V_{sd}$.  $I_K$ appears
almost linear to $V_{sd}$ at 480~mK, whereas nonlinearity becomes
prominent as $T$ decreases, reflecting the appearance of the Kondo
effect. The lower panel of Fig.~3(a) shows a plot of $S_K$ against
$I_K$. At any given $I_K$, $S_K$ increases with decreasing
$T$. Remarkably, $S_K$ is proportional to $|I_K|$ for all temperatures
and $F_K$ can be determined by numerical fitting to Eq.~(1). The
obtained $F_K$ is shown in Fig.~3(b) as a function of $T$. At $T \gtrsim
T_K = 0.70$~K, $F_K = 0.1 \pm 0.3$. As $T$ decreases, $F_K$ increases
and reaches $1.8 \pm 0.2$ at 130 mK. This suggests that $F_K$ clearly
indicates the evolution of the Kondo effect and that $F_K > 1$ indicates
the electron bunching due to Kondo correlation. This observation of the
evolution of the nonequilibrium Kondo state as characterized by the
growth of $F_K$ with decreasing temperature is the main result of the
present work.

Theoretically, $F_K$ is predicted to be
5/3~\cite{SelaPRL2006,GolubPRB2006,GogolinPRL2006,MoraPRL2008,MoraPRB2009_2,FujiiJPSJ2010},
which is close to the value obtained here. In a realistic experimental
situation, however, $F_K$ may be smaller owing to several
reasons. First, the asymmetry of the QD causes $F_K$ to be smaller,
i.e., $F_K = 5/3 - 8\delta^2/3$, at $T = 0$~\cite{MoraPRL2008}. Although
no theoretical results are available for finite $\delta^2$ at finite
$T$, $F_K = 1.2$ is expected for $\delta^2 = 1 -G_0/G_q =0.17$ (present
case) at $T = 0$. Second, the finite $U/\Gamma$ also makes $F_K$
smaller. In our case, $F_K$ is predicted to be $\sim 1.6$ instead of
$5/3$~\cite{SakanoFujiiOguri}.  Thirdly, finite temperature affects the
estimation of $F_K$~\cite{MoraPRL2008}; even for the symmetric case
($\delta^2=0$), the value of $5/3$ is reached only in a high bias region
around $eV_{sd}/k_B T \gg 10$~\cite{MoraPRL2008}. Thus, although the
experimental value is in the same range as the theoretical one, the
difference between them is not negligible as it indicates that there
exists a mechanism for enhancing the shot noise of our QD in addition to
the theoretical value.  The reason for this is not yet understood. One
possibility is that non-Kondo transport processes~\cite{GrobisPRL2008}
such as inelastic cotunneling and/or finite transport through other
levels in the QD contribute to the shot noise. For this finite transport
process, $U$ and $\Gamma$ are of the same order in the present case;
therefore, finite transport via the other adjacent levels, which are
irrelevant to the Kondo state but are strongly coupled with the leads,
might be possible. Even if the conductance of such transport is small,
the backscattering due to it would cause an enhancement of the shot
noise in the QD.

The above experimental result is reproducible in a wide parameter range
for the single-particle level position.  Figure~3(c) shows the obtained
$F_K$ with $T_K$ as a function of $V_g$. The universality of Kondo
physics predicts that $F_K$ should be constant regardless of the level
position at $T \ll T_K$; however, in our case, $F_K$ is found to be
weakly dependent on $V_g$ at 130~mK.  This fact might support the
aforementioned possible additional contribution to noise.  At 480~mK,
$F_K$ becomes almost constant at a value less than unity.  While $T_K$
does not drastically vary as a function of $V_g$, the evolution of the
Kondo effect is clearly visible in the case of a wide parameter range
when $T$ decreases below $T_K$.

Finally, the enhancement of the Fano factor is sometimes referred to as
``effective charge'' ($e^*$) that is larger than $e$, although such
$e^*$ deduced by the measurement of shot noise in the Kondo regime
should not be confused with an exotic charge like that in the fractional
quantum Hall
effect~\cite{SelaPRL2006,GogolinPRL2006,MoraPRL2008,MoraPRB2009_2}. In
this sense, an analysis based on the Fano factor would be preferable to
that based on $e^*$. Nevertheless, it is necessary to perform an
``effective charge'' analysis in order to compare our result with the
experimental result that $e^*/e \sim 5/3$, which was in agreement with
the theoretically predicted value~\cite{ZarchinPRB2008}. In Fig.~4(a),
the excess noise at $V_g = -0.740$~V, where the contribution of thermal
noise has been subtracted, is shown for 130, 300, and 480~mK.  We fit
these to $S_I^{ex} (V_{sd}) = 2e^* V_{sd} G (1-G/G_q)\left( \coth(x) -
1/x\right)$, as was done before~\cite{ZarchinPRB2008}.  The bias window
of this fitting is set to $|eV_{sd}| \ll k_B T_K$. As shown in
Figs.~4(a) and 4(b), $e^*/e = 2.9 \pm 0.2$ is obtained at 130~mK,
whereas $e^*/e$ is close to unity above $T_K = 0.70$~K. $e^*/e \sim 1$
was reported when the QD was tuned (by the gate voltage) to be outside
the Kondo regime~\cite{ZarchinPRB2008}.

\begin{figure}[tbp]
\center \includegraphics[width=.75\linewidth]{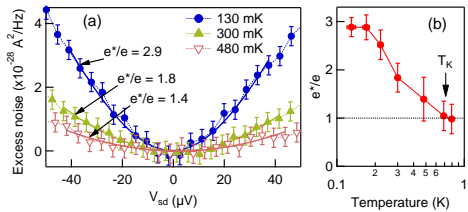}
\caption{(a) Excess current noise for $V_g = -0.740$~V at 130~mK,
300~mK, and 480~mK as a function of $V_{sd}$.  The solid curves are the
result of fitting with the obtained $e^*/e$ indicated.  (b) Obtained $e^*/e$ as a function of $T$.}
\end{figure}

It is naively expected that $e^*/e$ derived in this way equals $F_K$ at
$T = 0$~\cite{SelaPRL2006,GolubPRB2006}. In the Kondo system at finite
temperature, however, it is reasonable that $F_K$ and $e^*/e$ are
different from each other, because the procedures for determining them
are essentially different; the finite temperature that would affect the
distance from the Kondo fixed point is taken into account in obtaining
$F_K$. In contrast, to derive $e^*/e$, the system is treated as if it
would obey noninteracting physics. The present $e^*/e$ value of 2.9 at
130~mK is larger than the theoretical estimate of $5/3$. As discussed
above, $F_K$ (and hence $e^*/e$ at $T = 0$) is predicted to become
smaller than $5/3$ owing to several reasons. Therefore, both the present
result and the value close to $5/3$ reported in an asymmetric
QD~\cite{ZarchinPRB2008} support the suggestion that additional noise
sources enhance the shot noise in the Kondo QD.

To conclude, we successfully observed the evolution of the Kondo state
through the measurement of shot noise in a Kondo QD. With decreasing
temperature, the Fano factor increases and exceeds the value in the
noninteracting case, owing to two-particle scattering. The obtained
factors are larger than the theoretical factor, which may indicate that
the shot noise in a Kondo QD is larger than the theoretical prediction.
Further experimental effort, e.g., study of shot noise at the unitary
limit in a perfectly symmetric QD, is necessary to quantitatively
elucidate the nonequilibrium aspects of Kondo physics.

We are thankful to K. Kang, M. Hashisaka, S. Sasaki, and R. S. Deacon
for fruitful discussions. This work was partially supported by KAKENHI
and the Yamada Science Foundation.

\end{document}